\documentclass[aps,prapplied,reprint,superscriptaddress,nofootinbib,longbibliography]{revtex4-2}
\usepackage{amsmath,amssymb,bm,mathtools}
\usepackage{graphicx}
\usepackage[colorlinks=true,linkcolor=blue,citecolor=blue,urlcolor=blue]{hyperref}

\newcommand{\dd}{\mathrm{d}}

\newcommand{\phizero}{\phi_0}
\newcommand{\epsn}{\varepsilon_n}
\newcommand{\bfr}{\bm{r}}

\newcommand{\bfA}{\bm{A}}
\newcommand{\bfq}{\bm{q}}

\newcommand{\bfj}{\bm{j}}

\newcommand{\zhat}{\hat{\bm{z}}}
\newcommand{\yhat}{\hat{\bm{y}}}
\newcommand{\xhat}{\hat{\bm{x}}}

\newcommand{\Gibbs}{\mathcal{G}}
\newcommand{\Ns}{N_0}

\begin{document}
\raggedbottom

\title{Microscopic theory of the lower critical field in superconducting thin-film strips}

\author{Takayuki Kubo}
\email[]{kubotaka@post.kek.jp}
\affiliation{High Energy Accelerator Research Organization (KEK), Tsukuba, Ibaraki 305-0801, Japan}
\affiliation{The Graduate University for Advanced Studies (Sokendai), Hayama, Kanagawa 240-0193, Japan}

%\date{\today}

\begin{abstract}
The lower critical field \(B_{c1}\) of a narrow superconducting thin-film strip sets the thermodynamic scale for vortex-free operation in a perpendicular magnetic field.  
The standard Pearl--London estimate requires a phenomenological vortex-core cutoff, because the London
theory does not resolve the core.  
We formulate a microscopic theory for a dirty strip by solving the two-dimensional Usadel equations in the film plane, with the applied field included directly in the gauge-invariant momentum.  
Self-consistent vortex and Meissner solutions are computed at fixed field, and \(B_{c1}\) is obtained from their Gibbs-energy difference.  
The calculation resolves the vortex core and its finite-width deformation without introducing a cutoff.  
The resulting vortex self-energy is larger than the naive Pearl--London estimate and cannot, in general, be represented by a London logarithm with a single width-independent cutoff.  
The formulation applies at any \(T<T_c\) and provides a microscopic basis for predicting \(B_{c1}\) in superconducting nanostrips and related thin-film devices.
\end{abstract}

\maketitle

%%%%%%%%%%%%%
% Introduction
%%%%%%%%%%%%%

The lower critical field \(B_{c1}\) is the field above which a vortex becomes thermodynamically favorable relative to the vortex-free state.
In bulk superconductors, including those used in superconducting radio-frequency accelerator cavities, 
\(B_{c1}\) is a familiar thermodynamic scale for magnetic-flux penetration and field limitations~\cite{Kubo_2025}, with the superheating field setting the ideal limit.  
In narrow thin-film strips, which are used in many superconducting devices, 
\(B_{c1}\) has a more geometrical role: it sets the perpendicular-field scale above which a stable vortex can exist in the strip, and therefore gives a measure of the vortex-free field tolerance of the strip~\cite{Martinis, Bronson, Kuit}.

\begin{figure}[t]
\includegraphics[width=\columnwidth]{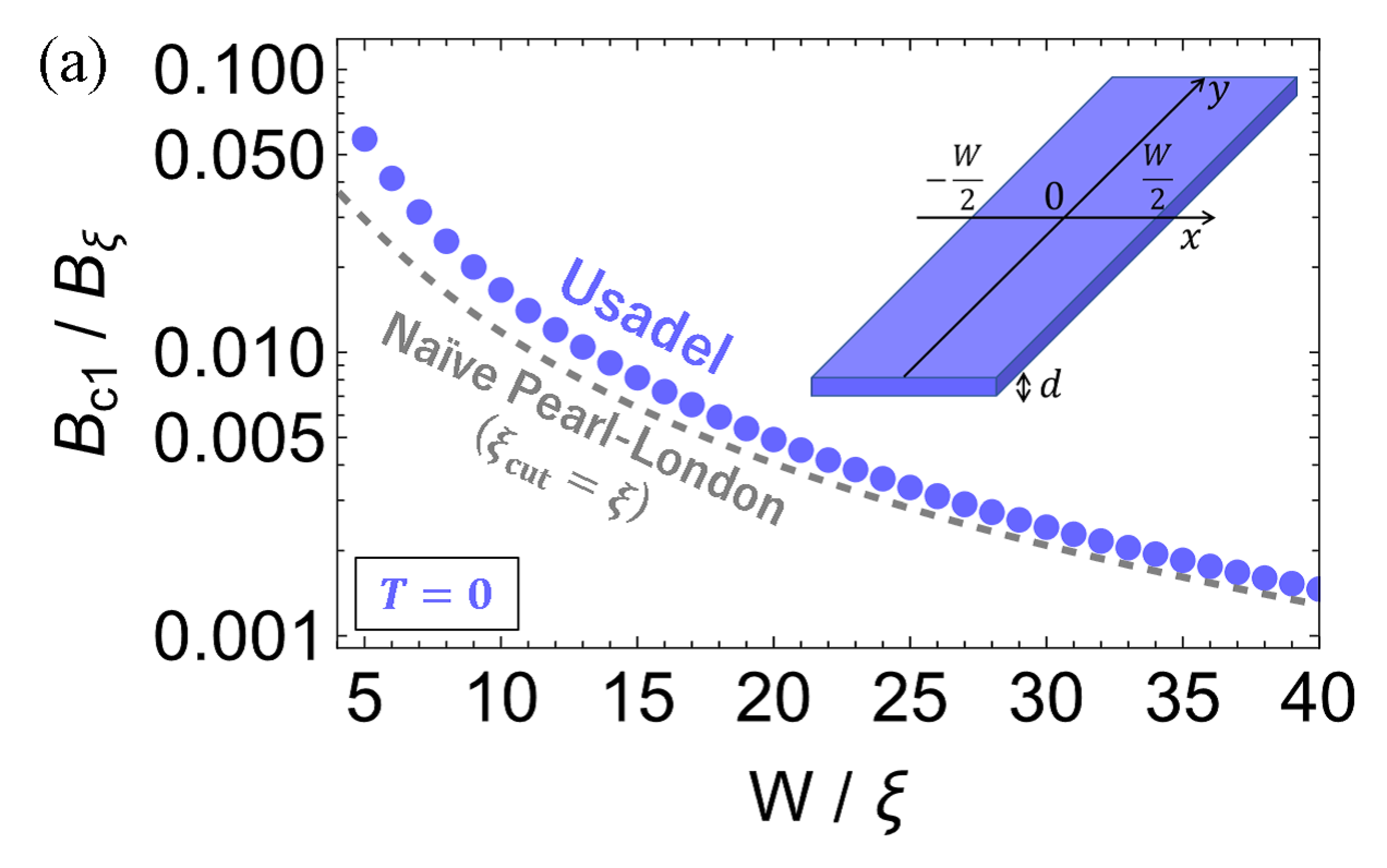}
\includegraphics[width=\columnwidth]{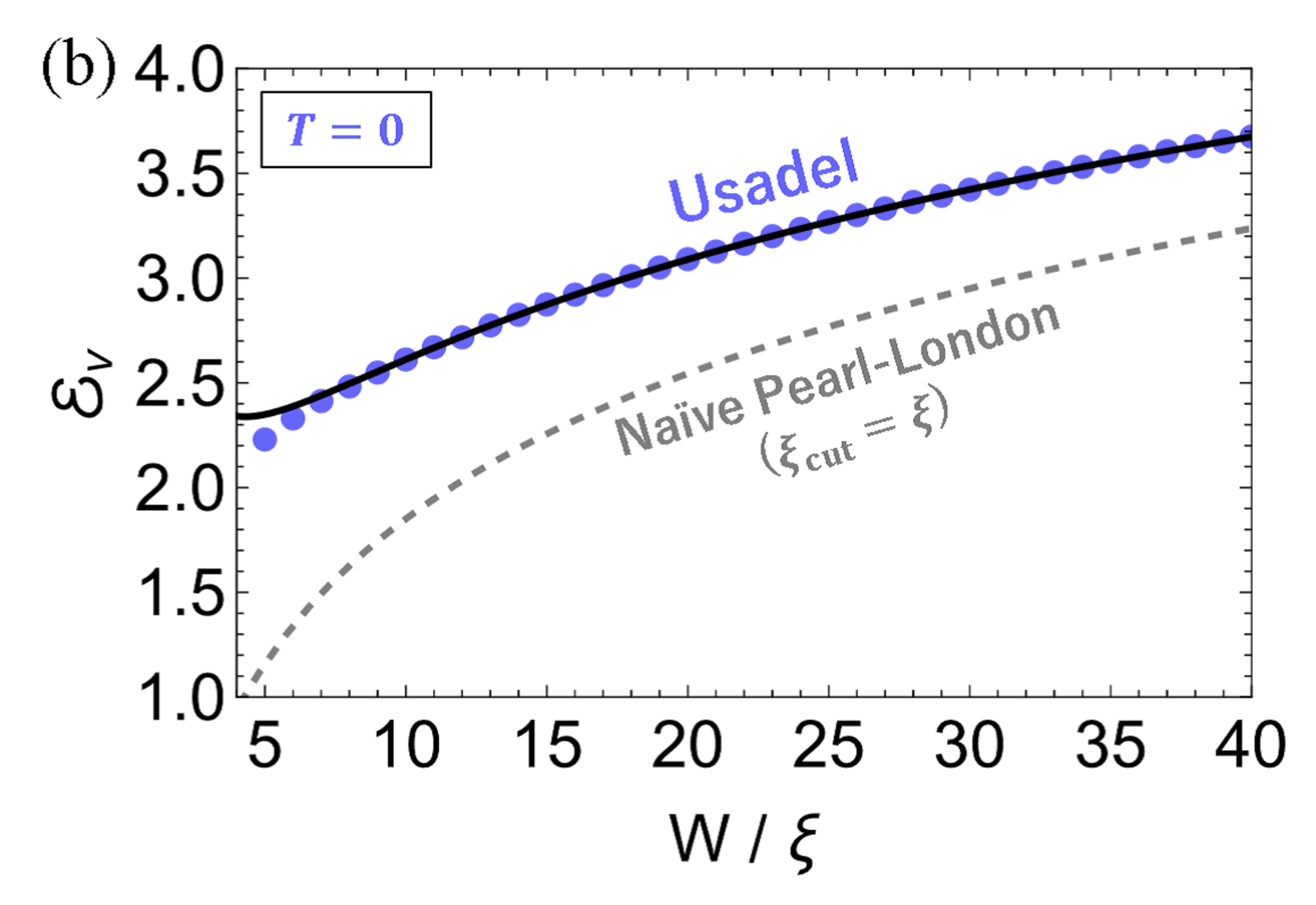}
\caption{
(a) Lower critical field \(B_{c1}\) of a narrow dirty superconducting strip at \(T=0\), 
normalized by \(B_\xi: =\phizero/\xi^2\), as a function of \(w:=W/\xi\).  
Blue symbols show the self-consistent two-dimensional Usadel results. 
The gray dashed curve is the naive Pearl-London estimate with \(\xi_{\rm cut}=\xi\).
(b) The same data expressed as the vortex self-energy factor \(\mathcal E_{\rm v}\), 
defined by \(B_{c1}=(2\phizero/\pi W^2)\mathcal E_{\rm v}\). 
The gray dashed curve is the naive Pearl-London estimate,
\(\mathcal E_{\rm v}^{\rm L}=\ln[2w/\pi]\).  
The black curve is a Pearl--London-type fit, 
\(\mathcal E_{\rm v}^{\rm fit}=\ln[2w/(\pi c_1)]+c_2/w\), with \(w=W/\xi\). 
Equivalently, this fit can be written in the Pearl--London form by introducing a width-dependent cutoff
\(\xi_{\rm cut}=c_1\xi\exp(-c_2/w)\).  
}
\label{fig1}
\end{figure}

The problem of determining \(B_{c1}\) in narrow superconducting thin-film strips has been studied for several decades, with the standard analytical treatment based on Pearl--London theory~\cite{Martinis, Bronson, Kuit, Pearl, Likharev, Kogan_1994, Maksimova, Clem_1998, Kogan_2020, Kogan_2021, Kubo_2023}. 
We consider a long strip extending along the \(y\) direction, with width \(W\) in the \(x\) direction and occupying \(-W/2<x<W/2\) [see the inset of Figure~\ref{fig1} (a)].  
The film thickness \(d\) is much smaller than the London penetration depth \(\lambda\).  
The applied field and the vortex line are along \(z\), and the vortex position in the film plane is denoted by \(\bfr_{\rm v}=(X,0)\).  
In the thin-strip limit \(d\ll\lambda\) and \(W\ll\Lambda=2\lambda^2/d\), 
where \(\Lambda\) is the Pearl length, magnetic screening by the strip is weak.
In this limit, Pearl-London theory gives \(B_{c1}^{\rm L}=(2\phizero/\pi W^2)\mathcal E_{\rm v}^{\rm L}\), where \(\mathcal E_{\rm v}^{\rm L}=\ln[2W/(\pi\xi_{\rm cut})]\). 
Here \(\mathcal E_{\rm v}\) is the dimensionless vortex self-energy factor, and \(\xi_{\rm cut}\) is a short-distance cutoff introduced to regularize the vortex-core singularity.

The factor \(\mathcal E_{\rm v}^{\rm L}\) contains both the long-range circulating-current energy and the short-distance core contribution.
Pearl-London theory describes the former, but it does not resolve the vortex core. 
Therefore, the short-distance cutoff \(\xi_{\rm cut}\) has to be introduced by hand.  
This cutoff is not a harmless microscopic detail.  
Since the core contribution is part of \(\mathcal E_{\rm v}^{\rm L}\), 
the predicted value of \(B_{c1}^{\rm L}\) depends on the assumed cutoff.  
Its temperature and material dependence must also be introduced through an assumed \(\xi_{\rm cut}(T)\), which London theory itself cannot determine.
Despite this long history, a quantitative determination of \(B_{c1}\) in narrow thin-film strips over the full superconducting temperature range has remained unresolved.

In this Letter, we remove this ambiguity by determining \(B_{c1}\) microscopically.  
We solve the two-dimensional dirty-limit Usadel equations~\cite{Usadel, Kopnin} in the film plane, with the applied field included directly in the gauge-invariant momentum.  
The spatial suppression of the order parameter in and around the vortex core is computed self-consistently,
and therefore we do not introduce a phenomenological core cutoff.  
For each applied field, we compute the Gibbs energy difference \(\Gibbs(X,B)\) between a state containing a vortex at position \(X\) and the vortex-free state at the same field.  
For the uniform symmetric strip, the lowest-energy vortex configuration is centered, and \(B_{c1}\) is obtained from \(\Gibbs(0,B_{c1})=0\).  
This gives a microscopic determination of \(B_{c1}\) in a superconducting thin-film strip without using a London core cutoff.

%%%%%%%%%
% Theory
%%%%%%%%%

We assume that the film is thin enough that \(\Delta\) and the Green functions are uniform across the thickness; the thickness \(d\) then enters only as an overall factor in the electronic free energy.
We parametrize the quasiclassical Matsubara Green functions as \(G_n=\cos\theta_n\) and \(F_n=\sin\theta_n\). 
Our convention and notations follow Refs.~\cite{Gurevich_Kubo, Kubo_2020, Kubo_2020_erratum}.
The corrections in the Erratum~\cite{Kubo_2020_erratum} do not affect the expressions used here. 
For positive Matsubara energies \(\epsn=2\pi k_B T(n+1/2)\), the self-consistent problem is
defined by the Usadel equation, the gap equation, and supercurrent
conservation:
\begin{eqnarray}
&&
\frac{\hbar D}{2}
\left(\nabla^2\theta_n-q^2\sin\theta_n\cos\theta_n\right)
=
\epsn\sin\theta_n-\Delta\cos\theta_n,
\label{eq:dim_usadel}
\\
&&
\Delta\ln\frac{T}{T_c}
=
2\pi k_B T\sum_{\epsn>0}
\left(\sin\theta_n-\frac{\Delta}{\epsn}\right),
\label{eq:dim_gap}
\\
&&
0=
\nabla\cdot\left[S(\bfr)\bfq(\bfr)\right],
\qquad
S(\bfr)=\sum_{\epsn>0}\sin^2\theta_n(\bfr).
\label{eq:dim_phase}
\end{eqnarray}
Here \(\Delta(\bfr)\) is the real order-parameter amplitude,
\(q=|\bfq|\), and \(D\) is the diffusivity. 
The supercurrent is $\bfj_s= -(2\pi\sigma_n k_B T/|e|) S(\bfr)\bfq(\bfr)$, 
where \(\sigma_n=2e^2N_0D\) is the normal-state conductivity, 
so Eq.~\eqref{eq:dim_phase} expresses local current conservation.
For the Meissner state and the state containing a single vortex,
respectively, we write the gauge-invariant momentum as
\begin{eqnarray}
\bfq_{\rm M}
&=&
\nabla\varphi_{\rm M}
+\frac{2\pi}{\phizero}Bx\,\yhat,
\label{eq:q_meissner}
\\
\bfq_{\rm v}
&=&
\nabla\varphi_{\rm v}
+\frac{2\pi}{\phizero}Bx\,\yhat
-2\pi\nabla\times (\psi_X\zhat),
\label{eq:q_vortex}
\end{eqnarray}
Here we choose the external vector potential $\bfA_{\rm ext}=Bx\,\yhat$. 
Since we consider a narrow strip with $W/\Lambda\ll1$, the magnetic screening of the applied field and the magnetic self-field of the vortex are negligible. 
Accordingly, the vector potential generated by the supercurrents gives only a small correction to $\bfA_{\rm ext}$ and is neglected. 
When $W$ becomes comparable to $\Lambda$, electromagnetic screening must be treated self-consistently.
For the Meissner state, $\varphi_{\rm M}=0$ by symmetry.  
For the vortex state, $\varphi_{\rm v}$ is determined by current conservation with zero net transport current along the strip, $d\int_{-W/2}^{W/2}j_{s,y}(x,y)\,dx=0$. 
For the centered vortex considered in the calculation of $B_{c1}$, this condition is automatically satisfied by reflection symmetry. 
For a vortex fixed at \(\bfr_{\rm v}=(X,0)\), the scalar Green function \(\psi_X\) is defined by
\(-\nabla^2\psi_X=\delta^{(2)}(\bfr-\bfr_{\rm v})\), with \(\psi_X(\pm W/2,y)=0\) and
\(\psi_X(x,y)\to0\) as \(|y|\to\infty\).  
In the numerical calculation, the \(y\) direction is truncated to a finite domain, 
and the length is chosen large enough that \(\mathcal E_{\rm v}\) is insensitive to it.
The last term in Eq.~\eqref{eq:q_vortex} is the singular vortex contribution; we denote it by
\(\bfq_X\equiv-2\pi\nabla\times(\psi_X\zhat)\).  
This representation satisfies the no-normal-flow condition at the strip edges.  
Although it has a stream-function form, \(\psi_X\) is not the stream function of the physical current, because the current also contains the spatially varying factor \(S(\bfr)\) and the remaining terms in \(\bfq_{\rm v}\).
With our sign convention, \(\oint_{\mathcal C_{\rm v}}\bfq_X\cdot\dd\bm\ell=-2\pi\), where
\(\mathcal C_{\rm v}\) is a small contour enclosing the vortex core.  
The minus sign selects the vortex orientation whose magnetic moment is parallel to the applied field \(B\zhat\), because the supercurrent is antiparallel to \(\bfq\).

At the side edges \(x=\pm W/2\), we denote the outward unit normal by \(\hat{\bm n}\), so that \(\hat{\bm n}=\pm\xhat\).  
We impose the insulating-boundary conditions
\begin{eqnarray}
\hat{\bm n}\cdot\nabla\theta_n=0,
\qquad
\hat{\bm n}\cdot\bfj_s=0.
\label{eq:side_boundary_conditions}
\end{eqnarray}
The second condition is equivalently \(\hat{\bm n}\cdot[S(\bfr)\bfq(\bfr)]=0\).  
These conditions state that neither the spectral Green-function current nor the physical
supercurrent flows out of the strip through the side edges.  
We do not include any additional surface pair-breaking term at the side boundary.
No independent boundary condition for \(\Delta\) is imposed, because the gap amplitude is determined locally by the self-consistency condition, Eq.~\eqref{eq:dim_gap}.
Along the $y$ direction, the calculation is performed on a finite domain $-L_y/2\le y\le L_y/2$. 
At the truncated ends, $\Delta$ and $\theta_n$ are fixed to their Meissner-state values at the same applied field, and $\varphi_{\rm v}=0$. 
We use $L_y\ge6W$ in the calculations reported below;
for $W=40\xi$, increasing $L_y$ from $240\xi$ to $280\xi$ changes $B_{c1}$ by less than $0.01\%$.

For each self-consistent solution, we evaluate the superconducting electronic free energy
\begin{eqnarray}
F_\alpha &=& d\times \int_\Omega\dd^2r\, \mathcal F[\Delta_\alpha,\{\theta_{\alpha n}\},\bfq_\alpha], \qquad (\alpha={\rm v},{\rm M}), \label{eq:F_alpha} \\
\frac{\mathcal F}{\Ns} &=& \Delta^2\ln\frac{T}{T_c} +2\pi k_B T\sum_{\epsn>0}\biggl[
\frac{\Delta^2}{\epsn}-2\Delta\sin\theta_n
\nonumber\\
&+& 2\epsn(1-\cos\theta_n)+\frac{\hbar D}{2} \bigl\{(\nabla\theta_n)^2+q^2\sin^2\theta_n\bigr\}
\biggr], 
\label{eq:free_energy_density_dimensional}
\end{eqnarray}
Here \(\Omega\) is the two-dimensional domain of the strip, \(d\) is the film thickness, and \(\Ns\) is the normal-state density of states used for normalization.  
The fields \(\Delta_\alpha(\bfr)\), \(\theta_{\alpha n}(\bfr)\), and \(\varphi_\alpha(\bfr)\) are the self-consistent fields, and \(\bfq_\alpha(\bfr)\) is the corresponding gauge-invariant momentum.

Since magnetic screening by the strip is neglected, the field energy of the externally imposed field is common to the vortex and Meissner solutions and cancels in the difference. 
The Gibbs energy difference relevant for \(B_{c1}\) is therefore
\begin{eqnarray}
\Gibbs(X,B)= F_{\rm v}(X,B)-F_{\rm M}(B).
\label{eq:G_main}
\end{eqnarray}
The applied field has not been discarded; it remains in the electronic free energies through \(\bfq_{\rm M}\) and \(\bfq_{\rm v}\).

The lower critical field is the field at which the lowest-energy vortex configuration first becomes degenerate with the Meissner state.  
For the uniform symmetric strip considered here, this configuration is centered at \(X=0\).
We therefore determine \(B_{c1}\) from
\begin{eqnarray}
\Gibbs(0,B_{c1})=0 . 
\label{eq:bc1_center_main}
\end{eqnarray}

We compute \(\Gibbs(X,B)\) directly from self-consistent vortex and Meissner solutions at the same applied field, obtained from Eqs.~\eqref{eq:dim_usadel}--\eqref{eq:dim_phase}.  
The physical inputs are \(T/T_c\) and \(W/\xi\). 
Here $\xi=\sqrt{\hbar D/(2\Delta_0)}$ is a widely used temperature-independent dirty-limit coherence length defined in terms of the zero-temperature gap $\Delta_0$, and $B_\xi:=\phizero/\xi^2$.  
It should be distinguished from the temperature-dependent Ginzburg--Landau coherence length, which near
$T_c$ behaves as $\xi_{\rm GL}(T)\simeq1.18\,\xi/\sqrt{1-T/T_c}$.
The self-consistent equations are solved on a nonuniform mesh refined near the vortex center.  
Depending on the strip width, we use grids from $61\times241$ to $161\times721$, with $L_y/\xi=72$--$240$.
The singular vortex phase winding is included analytically through $\bfq_X$, while $\Delta$ and $\theta_n$ are set to zero within a small numerical regularization region of radius $0.1\xi$ around the vortex center.  The coupled equations are iterated until the maximum changes in $\Delta$ and $\theta_n$ are below $5\times10^{-6}$ and $1\times10^{-5}$, respectively, and the relative residual of the phase equation is below $5\times10^{-5}$.  
Numerical convergence is checked by varying the mesh resolution, $L_y$, and the vortex-core regularization radius.  
The estimated numerical uncertainty in $B_{c1}$ is below $0.1\%$ for $T/T_c\le0.6$ and below $1\%$ at $T/T_c=0.8$.

%%%%%%%%%
% Results
%%%%%%%%%

We first focus on the zero-temperature limit.
Figure~\ref{fig1}(a) shows $B_{c1}/B_\xi$ as a function of $w:=W/\xi$ on a logarithmic vertical scale.
The vortex solution is obtained self-consistently, and no phenomenological core cutoff is introduced.  
The gray dashed curve is the naive Pearl-London estimate obtained by setting \(\xi_{\rm cut}=\xi\).  It lies systematically below the microscopic result.  
This difference is not very visible in Fig.~\ref{fig1}(a), because the overall magnitude of \(B_{c1}\) is
controlled mainly by the geometrical factor \(w^{-2}\), while the vortex core and material-dependent corrections enter only through the logarithmic self-energy factor.

To see the difference more clearly, we plot in Fig.~\ref{fig1}(b) the dimensionless vortex self-energy factor \(\mathcal E_{\rm v}\), defined by 
\begin{eqnarray}
B_{c1}=\frac{2\phizero}{\pi W^2} \mathcal E_{\rm v} .
\end{eqnarray}
The microscopic value exceeds the naive Pearl--London value by about \(14\%\) at \(w=40\) and \(34\%\) at \(w=12\).  More importantly, this difference cannot be described as a constant upward shift of the Pearl--London logarithm.  
A width-independent cutoff \(\xi_{\rm cut}=c_1\xi\) would only add a constant \(-\ln c_1\) to \(\mathcal E_{\rm v}^{\rm L}\), and is therefore not sufficient to fit the Usadel data. 
We therefore introduce a Pearl--London-like fitting function: 
\begin{eqnarray}
\mathcal E_{\rm v}^{\rm fit}=\ln \frac{2W}{\pi \xi_{\rm cut}},  \hspace{1cm}
\xi_{\rm cut}=c_1\xi e^{-c_2/w} .
\label{eq:fit}
\end{eqnarray}
Equivalently, \(\mathcal E_{\rm v}^{\rm fit}=\ln[2w/(\pi c_1)]+c_2/w\).  
The length \(c_1\xi\) gives the asymptotic microscopic core scale in the wide-strip limit \(W\gg\xi\), while \(c_2\) describes the leading finite-width correction that cannot be absorbed into a constant London cutoff.
A fit over $10\le w\le40$ gives $c_1=0.719$ and $c_2=4.31$.  
Based on the numerical convergence of the underlying $B_{c1}$ data, the uncertainties of the fitted coefficients are estimated to be of order $0.002$ and $0.03$, respectively.

\begin{figure}[t]
\includegraphics[width=\columnwidth]{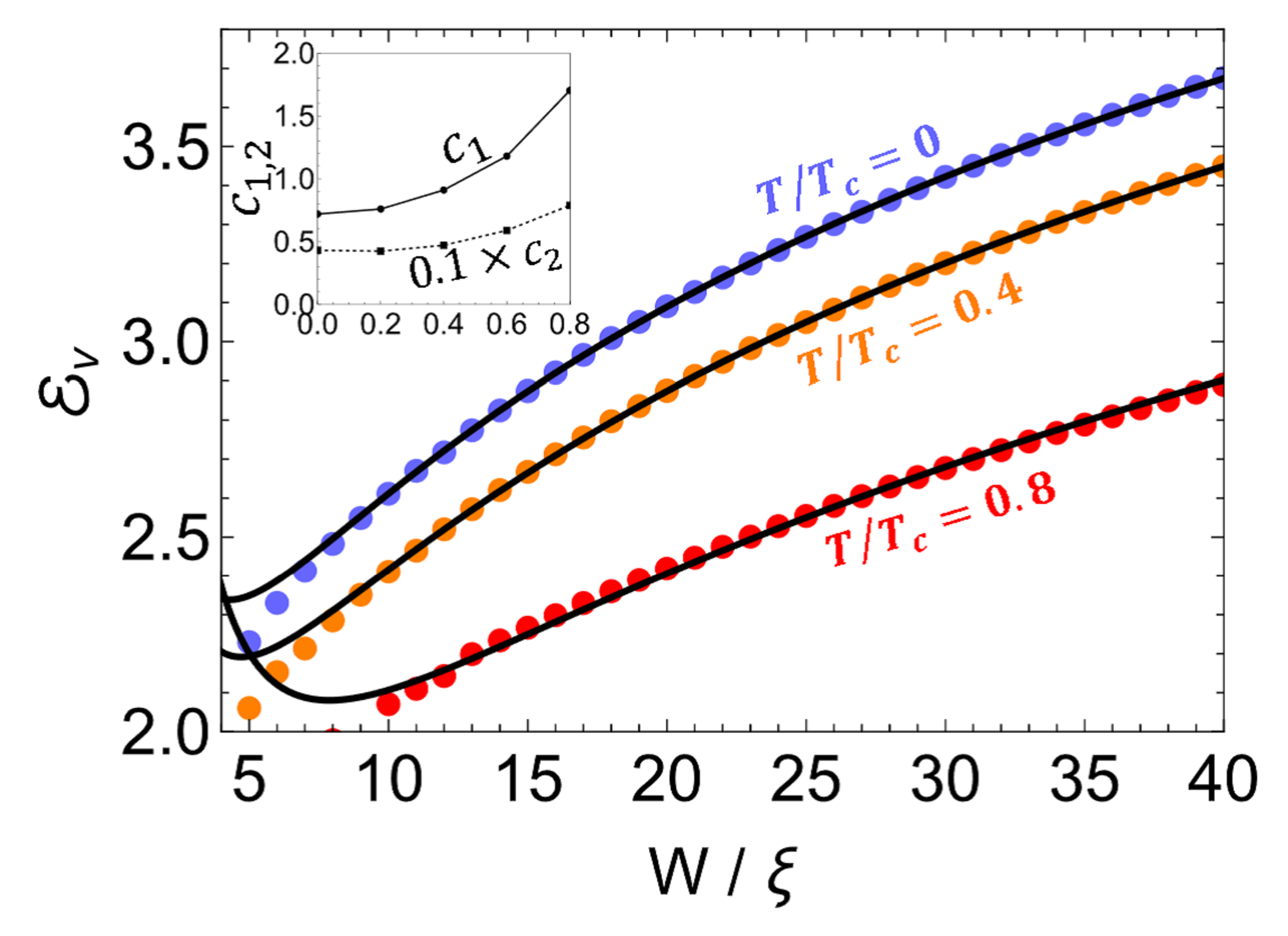}
\caption{
Finite-temperature behavior of the vortex self-energy factor \(\mathcal E_{\rm v}\).  
Symbols are the self-consistent two-dimensional Usadel results for \(T/T_c=0\), \(0.4\), and \(0.8\),
from top to bottom.  
The horizontal axis is \(w:=W/\xi\), where \(\xi=\sqrt{\hbar D/(2\Delta_0)}\) is the temperature-independent dirty-limit coherence length defined using the zero-temperature gap.  
The black curves are fits to the Pearl--London-type expression, Eq.~\eqref{eq:fit}. 
The inset shows the temperature ($T/T_c$) dependence of the fitted parameters \(c_1\) and \(0.1\times c_2\).
The fits use the data for \(10\le w\le40\).
}
\label{fig2}
\end{figure}

We next examine the finite-temperature behavior.  
Figure~\ref{fig2} shows representative results for \(\mathcal E_{\rm v}\) at several temperatures.  
In these calculations, the strip width is normalized by the temperature-independent dirty-limit coherence length \(\xi=\sqrt{\hbar D/(2\Delta_0)}\), defined using the zero-temperature gap \(\Delta_0\).  
Thus changing temperature does not redefine the strip width.  
Instead, the temperature dependence enters through the self-consistent gap, spectral functions, superfluid stiffness, and vortex-core structure.  
This is in contrast to the Pearl--London description, where the vortex core is not resolved and the temperature dependence must be introduced phenomenologically through a prescribed cutoff \(\xi_{\rm cut}(T)\), usually taken to be of order the coherence length.  
Pearl--London theory itself provides no microscopic rule for this cutoff.  
The finite-temperature Usadel results therefore provide a direct microscopic measure of how the effective core scale evolves with temperature.

The fitting form in Eq.~\eqref{eq:fit} also describes the finite-temperature data.  
The black curves overlaid on the microscopic results in Fig.~\ref{fig2} are fits to Eq.~\eqref{eq:fit}.  
The fitted parameters are $(c_1,c_2)=(0.719,4.31)$, $(0.910,4.70)$, and $(1.70,7.9)$ for $T/T_c=0$, $0.4$, and $0.8$, respectively.
For $T/T_c=0.2$ and $0.6$, which are not shown in Fig.~\ref{fig2}, the same analysis gives $(c_1,c_2)=(0.759,4.24)$ and $(1.18,5.88)$, respectively.  
From the numerical accuracy of the calculated $B_{c1}$, the uncertainties in $c_1$ and $c_2$ are estimated to be of order $0.002$ and $0.03$, respectively, for $T/T_c\le0.6$, and about $0.02$ and $0.2$ at $T/T_c=0.8$.

We have developed a two-dimensional Usadel theory for \(B_{c1}\) in a uniform narrow dirty superconducting strip.  
It resolves the vortex core self-consistently, without introducing the phenomenological Pearl--London cutoff, and applies at any \(T<T_c\).  
The result cannot, in general, be reduced to a London logarithm with a single width-independent cutoff.

In Pearl--London theory, the edge-barrier field \(B_s\), defined as the field at which the edge barrier for vortex entry disappears, is often calculated together with \(B_{c1}\) from the same analytical vortex
energy.  
Microscopically, however, \(B_s\) is not simply an energy-barrier calculation for a prescribed vortex configuration.  
It is the stability limit of the vortex-free state and should therefore be determined from the loss of local stability of this state, as in related stability analyses of superconducting states~\cite{Catelani,Transtrum,Gurevich_2023,Kubo_2026}.  
A microscopic calculation of $B_s$ based on the stability analysis of the Usadel free-energy functional has been given in Ref.~\cite{Kubo_Bs}.

The predicted width and temperature dependence of $B_{c1}$ could be tested by field-cooling narrow strips at different applied fields and directly imaging the trapped vortices, as in scanning Hall-probe measurements~\cite{Martinis}.  
Wide-field imaging with diamond NV centers~\cite{Kapur} and single-vortex-resolved magneto-optical imaging~\cite{Ooi_1, Ooi_2} provide alternative approaches for such measurements.

Finally, the method can be extended to nonuniform films.
Pearl--London theory predicts that spatial engineering of the Pearl length can enhance \(B_{c1}\)~\cite{Kubo_2023}.  
The corresponding problem can now be treated microscopically by allowing the Usadel parameters to depend on position.

\begin{acknowledgments}
This work was supported by JSPS KAKENHI under Grant Nos. JP25K01610, JP25K23386, JP26K03209, and JP26K00665. 
\end{acknowledgments}

\end{document}